\documentclass[floats, prd, eqnum, showpacs, nofootinbib, twocolumn,eqsecnum]{revtex4-1}

\usepackage{color,graphicx}
\usepackage{amsfonts}
\usepackage{amssymb}
\usepackage{comment}
\begin{document}

\title{Gravitational lensing by a Bronnikov-Kim wormhole under a weak-field approximation and in a strong deflection limit} 
\author{Naoki Tsukamoto${}^{1}$}\email{tsukamoto@rikkyo.ac.jp}

\affiliation{
${}^{1}$Department of General Science and Education, National Institute of Technology, Hachinohe College, Aomori 039-1192, Japan \\
}

\begin{abstract}
We consider gravitational lensing under a weak-field approximation and in a strong deflection limit 
by a Bronnikov-Kim wormhole with the same metric as the one of a wormhole which has been suggested in Einstein-Dirac-Maxwell theory. 
The metric approaches into the metric of an extreme charged Reissner-Nordstr\"{o}m black hole in a black hole limit 
and it becomes the metric of a spatial Schwarzschild wormhole in an ultrastatic limit.
In both of the black hole limit and the ultrastatic limit, 
the coefficient of a divergent term and the constant term of the deflection angle of a light in the strong deflection limit can be obtained exactly 
without expanding parameters of the spacetime. 
Interestingly, in the both limits to the black hole and the ultrastatic wormhole, 
we obtain exactly the same coefficient and constant term in the strong deflection limit. 
\end{abstract}
\maketitle

\section{Introduction}
Gravitational lensing under a weak-field approximation is a useful tool 
to find dark and massive objects~\cite{Schneider_Ehlers_Falco_1992,Schneider_Kochanek_Wambsganss_2006}.
Recently, the direct observations of gravitational waves from compact objects have been reported by LIGO Scientific and Virgo Collaborations~\cite{Abbott:2016blz}
and the shadow image of a supermassive black hole candidate at the center of galaxy M87 has been reported by Event Horizon Telescope Collaboration~\cite{Akiyama:2019cqa}. 
Gravitational lensing in a strong gravitational field by compact objects also can become an important tool to search them in future.
Gravitational lensing in a strong deflection limit~\cite{Bozza:2002zj} can be used to obtain
observables by a light ray reflected by a photon sphere, which is a sphere formed by unstable circular light orbits, 
around a compact object~\cite{Hagihara_1931,Darwin_1959, 
Atkinson_1965,Luminet_1979,Ohanian_1987,Nemiroff_1993,Frittelli_Kling_Newman_2000,Virbhadra_Ellis_2000,Bozza_Capozziello_Iovane_Scarpetta_2001,Bozza:2002zj,
Perlick:2003vg,Perlick_2004_Living_Rev,Virbhadra:2008ws,Bozza_2010,Tsukamoto:2016zdu,Shaikh:2019jfr,Shaikh:2019itn,Tsukamoto:2020uay,Tsukamoto:2020iez,Paul:2020ufc}.
The strong deflection limit analysis in the Schwarzschild spacetime was investigated by Bozza \textit{et al.}~\cite{Bozza_Capozziello_Iovane_Scarpetta_2001}
and its extensions and alternatives were investigated by several authors~\cite{Tsukamoto:2016zdu,Shaikh:2019jfr,Shaikh:2019itn,Tsukamoto:2020uay,Tsukamoto:2020iez,Paul:2020ufc,Bozza:2002zj,Bozza:2002af,Eiroa:2002mk,Petters:2002fa,Eiroa:2003jf,Eiroa:2004gh,Bozza:2004kq,Bozza:2005tg,Bozza:2006sn,Bozza:2006nm,Iyer:2006cn,Bozza:2007gt,Tsukamoto:2016qro,Ishihara:2016sfv,Tsukamoto:2016oca,Tsukamoto:2016jzh,Tsukamoto:2017edq,Hsieh:2021scb,Aldi:2016ntn,Tsukamoto:2020bjm,Hsieh:2021scb,Takizawa:2021gdp,Tsukamoto:2021caq,Aratore:2021usi}.

A wormhole is a hypothetical spacetime structure with nontrivial topology in general relativity~\cite{Visser_1995}
and it can be a black hole mimicker~\cite{Damour:2007ap,Lemos:2008cv,Tsukamoto:2019ihj}. 
It is known that energy conditions are violated at the throat of any asymptotically flat, static, and spherically symmetric wormhole at least
if general relativity without a cosmological constant is assumed~\cite{Morris:1988cz}.
Many passable wormholes were suggested after an Ellis-Bronnikov wormhole~\cite{Ellis_1973,Bronnikov_1973,Martinez:2020hjm} which is known as the earliest passable wormhole. 
In 2001, Dadhich \textit{et al.} considered wormhole metrics with a vanishing Ricci scalar~\cite{Dadhich:2001fu}
and then Bronnikov and Kim also suggested wormhole metrics with a vanishing Ricci scalar in a braneworld scenario~\cite{Bronnikov:2002rn,Bronnikov:2003gx}.
Gravitational lensing by a spatial Schwarzschild wormhole with the vanishing Ricci scalar~\cite{Dadhich:2001fu} 
in the strong deflection limit~\cite{Bozza:2002zj} 
has been investigated in Ref.~\cite{Tsukamoto:2016zdu}.

Recently, a wormhole filled with massless and neutral fermions in Einstein-Dirac-Maxwell theory was obtained numerically~\cite{Blazquez-Salcedo:2020czn}
and the existence of a thin shell at the wormhole throat was discussed in Refs.~\cite{Bolokhov:2021fil,Konoplya:2021hsm,Blazquez-Salcedo:2021udn,Danielson:2021aor}.
Its metric corresponds with the Bronnikov-Kim wormhole as a simple and analytical metric case~\cite{Bronnikov:2002rn,Bolokhov:2021fil,Blazquez-Salcedo:2021udn}.
Its shadow size also was investigated by Bronnikov \textit{et al.}~\cite{Bronnikov:2021liv}
and the orbit of a star around the wormhole was investigated by Jusufi \textit{et al.}~\cite{Jusufi:2021lei}. 
Moreover, Churilova \textit{et al.}~\cite{Churilova:2021tgn} discussed the shadow 
or the apparent size of a photon sphere of an asymmetric wormhole~\cite{Konoplya:2021hsm} in a small asymmetry case. 

In this paper, 
we consider gravitational lensing under the weak-field approximation and in the strong deflection limit
by the Bronnikov-Kim wormhole~\cite{Bronnikov:2002rn,Bronnikov:2003gx} 
with the same metric as the one of the wormhole which was suggested in Einstein-Dirac-Maxwell theory~\cite{Blazquez-Salcedo:2020czn}.
The metric approaches into the metric of an extreme charged Reissner-Nordstr\"{o}m black hole in a black hole limit 
and it approaches into the metric of the spatial Schwarzschild wormhole~\cite{Dadhich:2001fu} in an ultrastatic limit.

This paper is organized as follows. 
The deflection angle of a ray in the Bronnikov-Kim wormhole spacetime is obtained in Sec.~II, 
a lens equation is introduced in Sec.~III,
and observables under a weak-field approximation are obtained in Sec.~IV.
We review a Schwarzschild spacetime as a reference in Sec.~V.
The deflection angle and observables by the Bronnikov-Kim wormhole 
in the strong deflection limit are investigated in Sec. VI 
and then we discuss and conclude our results in Sec. VII.
We investigate an Arnowitt-Deser-Misner(ADM) mass in the Appendix.
In this paper, we use the units in which the light speed and Newton's constant are unity.

\section{Deflection angle in the Bronnikov-Kim wormhole spacetime}
The line element of a Bronnikov-Kim wormhole spacetime~\cite{Bronnikov:2002rn,Bronnikov:2003gx} is given by
\begin{equation}\label{eq:le}
ds^{2}
=-A(r)dt^{2} +B(r)dr^{2}+r^{2}(d\vartheta^{2}+\sin^{2}\vartheta d\varphi^{2}),
\end{equation}
where $A(r)$ and $B(r)$ are defined by
\begin{eqnarray}
A(r)\equiv \left( 1-\frac{M}{r} \right)^2
\end{eqnarray}
and 
\begin{eqnarray}
B(r) 
\equiv \frac{1}{\left( 1-\frac{r_+}{r} \right) \left(1-\frac{r_-}{r} \right)}
= \frac{1}{1-\frac{2Q^2}{Mr}+\frac{Q^2}{r^2}},
\end{eqnarray}
respectively,
where $M$ and $Q$ are positive constants which 
hold $0<M<Q$, and where $r_\pm$ is defined by
\begin{equation}
r_{\pm}\equiv \frac{Q^2}{M} \pm \sqrt{\frac{Q^4}{M^2}-Q^2}.
\end{equation}
There is a wormhole throat at $r=r_+$.
Note that $M/2<r_- < M < r_+$ as shown in Fig.~1.
In a black hole limit $Q\rightarrow M$, 
the metric approaches into the metric of the extreme charged Reissner-Nordstr\"{o}m black hole spacetime and we obtain $r_\pm=M=Q$.
There are time translational and  axial Killing vectors $t^{\mu}\partial_{\mu}=\partial_{t}$ and $\phi^{\mu}\partial_{\mu}=\partial_{\phi}$,
since the spacetime has stationarity and axisymmetry, respectively.
If the norm of the time translational Killing vector is a constant, 
the spacetime is called ultrastatic spacetime.
Notice that a relation $M=2Q^2r_+/(Q^2+r_+^2)$.
As shown in the Appendix, an ADM mass $m$ is given by 
\begin{equation}
m=\frac{Q^2}{M}.
\end{equation}
If we consider an ultrastatic limit $M \rightarrow 0$ under a fixed ADM mass $m$,
we obtain the metric of a spatial Schwarzschild wormhole~\cite{Dadhich:2001fu,Tsukamoto:2016zdu} in the following form:
\begin{eqnarray}
ds^{2}
=-dt^{2} +\frac{dr^{2}}{1-\frac{2m}{r}}+r^{2}(d\vartheta^{2}+\sin^{2}\vartheta d\varphi^{2}),
\end{eqnarray}
which has a vanishing Ricci scalar while it has nonzero components of Ricci tensors.
We can assume $\vartheta=\pi/2$ without loss of generality.
\begin{figure}[htbp]
\begin{center}
\includegraphics[width=85mm]{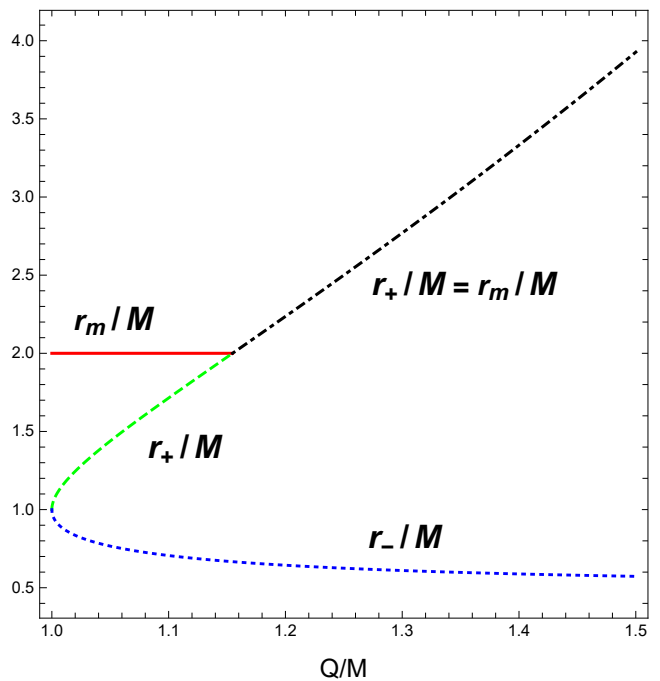}
\end{center}
\caption{The reduced radial coordinates of a photon sphere $r_\mathrm{m}/M$, 
throat $r_+/M$, 
and $r_-/M$ for given $Q/M$.
A (red) solid line,
a (green) dashed curve, 
a (blue) dotted curve,  
and a (black) dash-dotted curve 
denote $r_\mathrm{m}/M$ for $1 < Q/M \leq 2/\sqrt{3}$,
$r_{+}/M$ for $1 < Q/M \leq  2/\sqrt{3}$,
$r_{-}/M$ for $1 < Q/M$,
and $r_+/M=r_\mathrm{m}/M$ for $2/\sqrt{3} \leq Q/M$, respectively.
}
\label{fig:rq}
\end{figure}

From 
$k^{\mu}k_{\mu}=0$, where 
$k^{\mu}\equiv \dot{x}^{\mu}$ is a wave number vector and the overdot denotes a differentiation with respect to an affine parameter,
the trajectory of a ray is given by 
\begin{equation}\label{eq:locus}
-A\dot{t}^{2}+B\dot{r}^{2}+r^{2}\dot{\varphi}^{2}=0
\end{equation}
and it can be expressed by 
$\dot{r}^{2}+V(r)/E^2=0$,
where $V(r)$ is an effective potential defined by
\begin{equation}
V(r)\equiv \frac{1}{Br^2} \left( b^2 -\frac{r^2}{A} \right),
\end{equation}
where $b\equiv L/E$ is the impact parameter of the ray and 
$E\equiv -g_{\mu\nu}t^{\mu}k^{\nu}=A\dot{t}$, and $L\equiv g_{\mu\nu}\phi^{\mu}k^{\nu}=r^{2}\dot{\varphi}$
are the conserved energy and angular momentum of the light ray, respectively.
We assume that $L$ and $b$ are positive unless we focus on negative ones.
We assume that the closest distance of the light ray is at $r=r_0$. 
Note that $V_0\equiv V(r_0)=0$ holds. 
Here and hereinafter, any function with the subscript $0$ denotes the function at the closest distance $r=r_0$.   
Light rays can exist in a region for $V(r) \leq 0$.
For $1<Q/M<2/\sqrt{3}$, 
a ray with $b=b_\mathrm{m}=4M$ has $V_\mathrm{m}=V_\mathrm{m}^\prime=0$ and $V_\mathrm{m}^{\prime \prime}<0$ on the photon sphere at $r=r_{\mathrm{m}}=2M$ 
and a ray with $b=r_+^2/(r_+-M)$ has $V(r_+)=V^\prime(r_+)=0$ and $V^{\prime \prime}(r_+)>0$ 
on the throat acting as an antiphoton sphere,
which is a sphere formed by stable circular light orbits, at $r=r_{+}$.\footnote{%
It is a concern that stable circular light orbits may cause instability of ultracompact objects
because of the slow decay of linear waves~\cite{Keir:2014oka,Cardoso:2014sna,Cunha:2017qtt}.}
Here and hereinafter, the prime denotes the differentiation with respect to the radial coordinate $r$ or the closest distance $r_0$ and 
functions with the subscript m denote the functions at $r=r_\mathrm{m}$ or $r_0=r_\mathrm{m}$. The effective potentials are shown in Fig.~2.
For $Q/M>2/\sqrt{3}$, a ray with $b=b_\mathrm{m}=r_+^2/(r_+-M)$ has 
$V_\mathrm{m}=V_\mathrm{m}^\prime=0$ and $V_\mathrm{m}^{\prime \prime}<0$ on the throat which works as the photon sphere 
at $r=r_+=r_\mathrm{m}$ as shown Fig.~3. 
For a marginal case $Q/M=2/\sqrt{3}\sim 1.1547$, a ray with $b=b_\mathrm{m}=4M$ has 
$V_\mathrm{m}=V_\mathrm{m}^\prime=V_\mathrm{m}^{\prime \prime}=0$ and $V_\mathrm{m}^{\prime \prime \prime}=-4/M^3<0$ on the throat working as a marginally unstable photon sphere
at $r=r_\mathrm{m}=r_+=2M$.
A light ray falls into the throat at $r=r_+$ for $b<b_\mathrm{m}$,
it rotates around the photon sphere at $r=r_\mathrm{m}$ infinite times for $b=b_\mathrm{m}$,
and it is reflected by the throat for $b>b_\mathrm{m}$.
In this paper, we concentrate on light rays in the scattered case with $b>b_\mathrm{m}$ and gravitational lensing in a usual lens configuration.
\begin{figure}[htbp]
\begin{center}
\includegraphics[width=85mm]{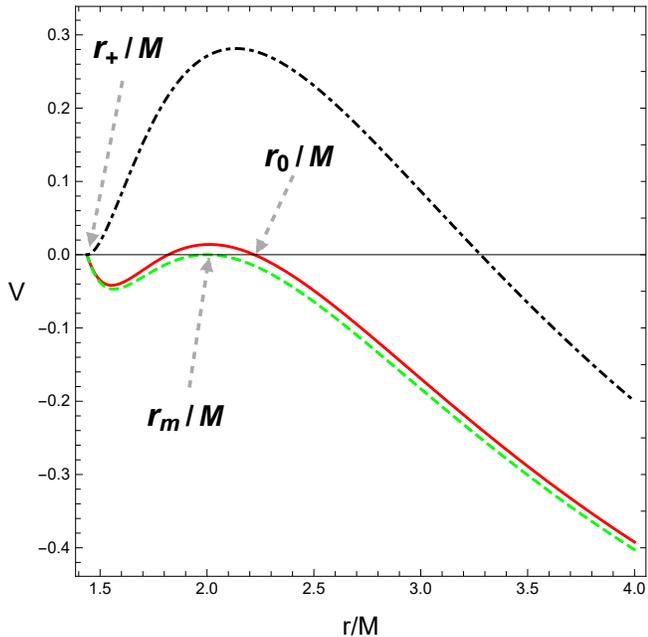}
\end{center}
\caption{Effective potentials~$V$ with~$Q/M=1.05$.
Solid (red), dashed (green), dot-dashed (black) curves denote effective potentials~$V$ 
with $b/M=1.01 b_\mathrm{m}/M=4.04$, $b/M=b_\mathrm{m}/M=4.00$, and $b/M=r_+^2/(r_+ -M)=4.72$, respectively.
}
\label{fig:P1}
\end{figure}
\begin{figure}[htbp]
\begin{center}
\includegraphics[width=85mm]{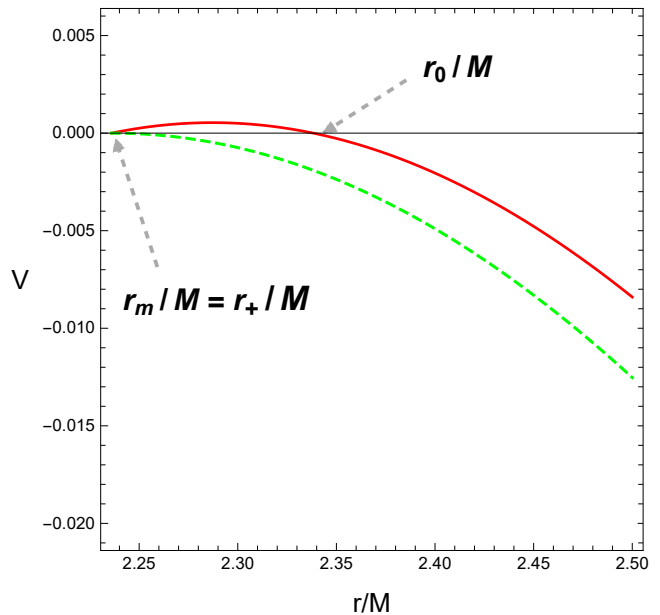}
\end{center}
\caption{Effective potentials~$V$ with~$Q/M=1.2$.
Solid (red) and dashed (green) curves denote effective potentials~$V$ 
with $b/M=1.01 b_\mathrm{m}/M=4.09$ and $b/m=b_\mathrm{m}/M=4.05$, respectively.
}
\label{fig:P2}
\end{figure}

From Eq.~(\ref{eq:locus}), we obtain 
\begin{equation}\label{eq:locus2}
A_0\dot{t}_0^{2}=r_0^{2}\dot{\varphi}^{2}_0
\end{equation}
and the impact parameter of the light can be expressed by, as a function of $r_0$,
\begin{equation}\label{eq:b0}
b(r_0)=\frac{L}{E}=\frac{r_0^2 \dot{\varphi}_0}{A_0\dot{t}_0}=\frac{r_0}{\sqrt{A_0}}=\frac{r_0^2}{r_0-M}.
\end{equation}
From the equation of the trajectory of the ray~(\ref{eq:locus}),
the deflection angle of the ray is given by
\begin{equation}
\alpha=I(r_0)-\pi,
\end{equation}
where $I(r_0)$ is defined as
\begin{equation}
I(r_0)\equiv 2\int^{r_0}_\infty \frac{b dr}{r^2\sqrt{-V(r)}}.
\end{equation}

\section{Lens equation}
We assume that an observer O and a source S are on the same side of a throat.
A light ray with an impact parameter $b$ is emitted by S with a source angle $\phi$,
it is reflected with a deflection angle $\alpha$ by a wormhole as a lens L, 
and it is observed by O as an image I with an image angle $\theta$.
We assume small angles~$\bar{\alpha}\ll 1$, $\theta=b/D_{\mathrm{ol}}\ll 1$, and $\phi \ll 1$,
where $D_{\mathrm{ol}}$ is a distance between O and L 
and $\bar{\alpha}$ is an effective deflection angle of the light ray defined by
\begin{equation}
\bar{\alpha} \equiv \alpha \quad  (\mathrm{mod} \quad  2\pi).
\end{equation}
We introduce the winding number $N$ of the light, and we can express the deflection angle as
\begin{equation}\label{eq:defn}
\alpha=\bar{\alpha}+ 2\pi N.
\end{equation}
\begin{figure}[htbp]
\begin{center}
\includegraphics[width=85mm]{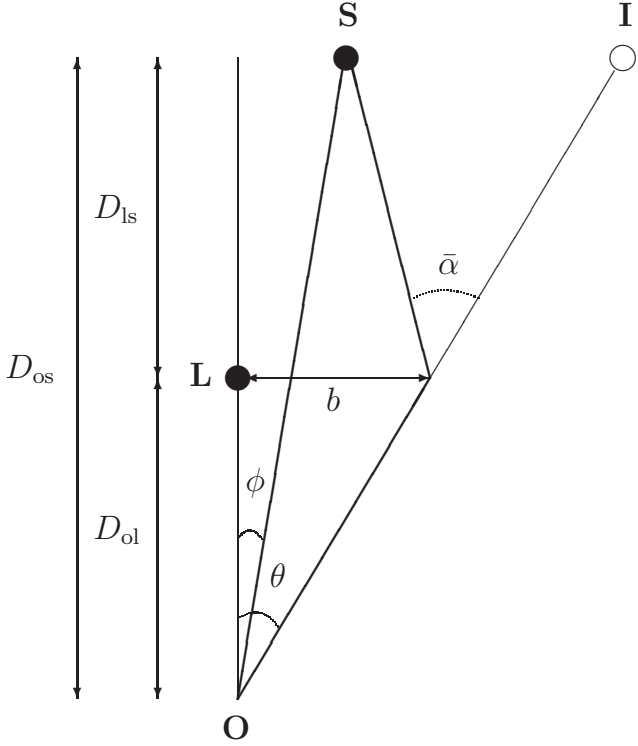}
\end{center}
\caption{Lens configuration. A source S with a source angle $\phi$ emits a light ray with an impact parameter $b$.  
It is reflected by a lens L with an effective deflection angle $\bar{\alpha}$ on a lens plane. 
An observer O sees it as an image I with an image angle $\theta$. 
Note that S and O are the same side of a wormhole throat.
The distances between O and S, between L and S, and between O and L are denoted by
$D_{\mathrm{os}}$, $D_{\mathrm{ls}}$, and $D_{\mathrm{ol}}$, respectively.}
\label{fig:configuration}
\end{figure}
A small-angle lens equation~\cite{Bozza:2008ev} is obtained from the lens configuration in Fig.~\ref{fig:configuration} as 
\begin{equation}\label{eq:lens}
D_\mathrm{ls} \bar{\alpha}=D_\mathrm{os} ( \theta-\phi ),
\end{equation}
where $D_{\mathrm{ls}}$ and $D_{\mathrm{os}}=D_{\mathrm{ol}}+D_{\mathrm{ls}}$ are 
the distances between L and S and between O and S, respectively.

\section{Gravitational lensing under a weak-field approximation}
Under a weak-field approximation $r_\mathrm{m} \ll r_0$ or $b_\mathrm{m} \ll b$,
since $M \ll r$ and $Q^2/M \ll r$ must be satisfied, 
the line element~(\ref{eq:le}) can be expanded as  
\begin{eqnarray}
ds^{2}
&\sim&-\left(1-\frac{2M}{r} \right)dt^{2} +\left(1+\frac{2\gamma M}{r} \right) dr^{2} \nonumber\\
&&+r^{2}(d\vartheta^{2}+\sin^{2}\vartheta d\varphi^{2}) \nonumber\\
&=&-\left(1-\frac{2M}{r} \right)dt^{2} +\left(1+\frac{2Q^2}{Mr} \right) dr^{2} \nonumber\\
&&+r^{2}(d\vartheta^{2}+\sin^{2}\vartheta d\varphi^{2}), 
\end{eqnarray}
where $\gamma$ is given by $\gamma \equiv Q^2/M^2$.
From Eq. (8.5.8) in Ref.~\cite{Weinberg:1972kfs} and $r_0\sim b$, 
the deflection angle under the weak-field approximation is given by  
\begin{equation}\label{eq:weakal}
\alpha
\sim \frac{4M}{b} \left( \frac{1+\gamma}{2} \right)
= \frac{2}{b} \left( M + \frac{Q^2}{M} \right) = \frac{2\left( M + m \right)}{b}. 
\end{equation}
Note that the winding number $N$ vanishes under the weak-field approximation.

From Eqs.~(\ref{eq:defn}) and $(\ref{eq:weakal})$, $N=0$, and $\theta=b/D_{\mathrm{ol}}$,
the lens equation~(\ref{eq:lens}) can be written as
\begin{equation}\label{eq:lens2}
\theta^2-\phi \theta=\theta_{\mathrm{E}0}^2,
\end{equation}
where $\theta_{\mathrm{E}0}$ is defined as
\begin{equation}
\theta_{\mathrm{E}0}\equiv \sqrt{\frac{2(M+m)D_{\mathrm{ls}}}{D_{\mathrm{os}}D_{\mathrm{ol}}}}.
\end{equation}
The positive and negative solutions of the lens equation~(\ref{eq:lens2}) are obtained 
as $\theta=\theta_{0 \pm}=\theta_{0 \pm}(\phi)$, where 
\begin{equation}
\theta_{0 \pm} \equiv \frac{1}{2} \left( \phi \pm \sqrt{\phi^2+4\theta_{\mathrm{E}0}^2} \right).
\end{equation}
Here and hereinafter, the upper (lower) sign is chosen for the positive (negative) image angle with a positive (negative) impact parameter.
In a perfect-aligned case with $\phi=0$, a ring-shaped image, which is called an Einstein ring, 
can be observed.
The radius of the Einstein ring is given by
\begin{equation}
\theta=\theta_{0+}(0) = -\theta_{0-}(0) = \theta_{\mathrm{E}0}.
\end{equation}

The magnifications for the images are given by
\begin{eqnarray}
\mu_{0 \pm}
&\equiv& \frac{\theta_{0 \pm}}{\phi} \frac{d\theta_{0 \pm}}{d\phi} \nonumber\\
&=& \frac{1}{4} \left( 2\pm \frac{\phi}{\sqrt{\phi^2+4\theta_{\mathrm{E}0}^2}} \pm \frac{\sqrt{\phi^2+4\theta_{\mathrm{E}0}^2}}{\phi} \right) 
\end{eqnarray}
and its total magnification becomes 
\begin{eqnarray}
\mu_{0\mathrm{tot}}
&\equiv& \left| \mu_{0+} \right|+ \left| \mu_{0-} \right| \nonumber\\
&=& \frac{1}{2} \left( \frac{\left|\phi\right|}{\sqrt{\phi^2+4\theta_{\mathrm{E}0}^2}} + \frac{\sqrt{\phi^2+4\theta_{\mathrm{E}0}^2}}{\left|\phi\right|} \right). 
\end{eqnarray}

\section{Gravitational lensing in a strong deflection limit}

\subsection{Deflection angle in a strong deflection limit}
We investigate the deflection angle of a ray in a strong deflection limit $b\rightarrow b_\mathrm{m}$ in the following form:
\begin{eqnarray}
\alpha 
&=&-\bar{a} \log \left( \frac{b}{b_\mathrm{m}}-1 \right) +\bar{b} \nonumber\\
&&+O \left( \left( \frac{b}{b_\mathrm{m}}-1 \right) \log \left( \frac{b}{b_\mathrm{m}}-1 \right) \right),
\end{eqnarray}
where $\bar{a}$ and $\bar{b}$ are obtained from the line element.~\footnote{ %
We should read the order of a following term $O(b-b_\mathrm{m})$ in Ref.~\cite{Bozza:2002zj} 
as $O \left( \left( b/b_\mathrm{m}-1 \right) \log \left( b/b_\mathrm{m}-1 \right) \right)$ 
as discussed in Refs.~\cite{Iyer:2006cn,Tsukamoto:2016qro,Tsukamoto:2016jzh}. }

\subsubsection{For $1<Q/M<2/\sqrt{3}$}
In the case of $1<Q/M<2/\sqrt{3}$, 
the photon sphere of a light ray with $b=b_\mathrm{m}=4M$ is at $r=r_{\mathrm{m}}=2M$.
We introduce a variable~\cite{Tsukamoto:2016jzh}
\begin{equation}
z\equiv 1-\frac{r_0}{r}
\end{equation}
and we rewrite $I(r_0)$ as 
\begin{equation}
I(r_0)=\int^{1}_0 R(z,r_0) F(z,r_0) dz,
\end{equation}
where $R(z,r_0)$ is given by
\begin{equation}
R(z,r_0)\equiv \frac{2 \sqrt{M} r_0}{ \sqrt{M\left[r_0^2+Q^2(1-z)^2\right]+2Q^2 r_0 (z-1)}}
\end{equation}
and $F(z,r_0)$ is defined by
\begin{equation}
F(z,r_0)\equiv \frac{1}{\sqrt{f(z,r_0)}},
\end{equation}
where $f(z,r_0)$ is given by
\begin{equation}
f(z,r_0)\equiv \frac{r_0^4}{b^2(r_0)\left[r_0+M(z-1)\right]^2}-(1-z)^2.
\end{equation}
We expand $f(z,r_0)$ around $z=0$ and obtain 
\begin{equation}
f(z,r_0)= c_1(r_0)z + c_2(r_0)z^2+ O(z^3),
\end{equation}
where $c_1(r_0)$ and $c_2(r_0)$ are given by 
\begin{eqnarray}
c_1(r_0)\equiv \frac{2(2M-r_0)}{M-r_0}
\end{eqnarray}
and
\begin{eqnarray}
c_2(r_0)\equiv \frac{2M^2+2Mr_0-r_0^2}{(M-r_0)^2},
\end{eqnarray}
respectively.
Here, we have used Eq.~(\ref{eq:b0}).
From $c_{1\mathrm{m}}\equiv c_1(r_\mathrm{m})=0$ and $c_{2\mathrm{m}}\equiv c_2(r_\mathrm{m})=2$, $F(z,r_0)$ diverges 
and the leading order of the divergence is $z^{-1}$ in the strong deflection limit $r_0 \rightarrow r_\mathrm{m}$.

We expand $c_{1}(r_0)$ and $b(r_0)$ around $r_0=r_\mathrm{m}$ as 
\begin{eqnarray}\label{eq:cs}
c_{1}(r_0)= c_{1\mathrm{m}}^\prime (r_0-r_\mathrm{m})+O\left(( r_0-r_\mathrm{m} )^2\right), 
\end{eqnarray}
where $c_{1\mathrm{m}}^\prime=2/M$,
and 
\begin{equation}\label{eq:bs}
b(r_0)= b_\mathrm{m} + \frac{1}{2} b^{\prime \prime}_{\mathrm{m}} (r_0-r_\mathrm{m})^2 +O\left(( r_0-r_\mathrm{m} )^3\right), 
\end{equation}
where $b^{\prime \prime}_{\mathrm{m}}=2/M$ since we use them later.

We separate $I$ into a divergent part $I_D$ and a regular part $I_R$.
The divergent part is defined by 
\begin{equation}
I_D=\int^{1}_0 R(0,r_\mathrm{m}) F_D(z,r_0) dz,
\end{equation}
where $R(0,r_\mathrm{m})$ is given by
\begin{equation}\label{eq:R0}
R(0,r_\mathrm{m})=\frac{4M}{\sqrt{4M^2-3Q^2}}
\end{equation}
and $F_D(z,r_0)$ is defined by
\begin{equation}\label{eq:fD}
F_D(z,r_0)\equiv \frac{1}{\sqrt{c_1(r_0) z+ c_2(r_0) z^2}}.
\end{equation}
We can integrate $I_D$ as~\cite{Bozza:2002zj,Tsukamoto:2016jzh}
\begin{equation}\label{eq:ID}
I_D = \frac{2R(0,r_\mathrm{m})}{\sqrt{c_2(r_0)}} \log \frac{\sqrt{c_2(r_0)}+\sqrt{c_1(r_0)+c_2(r_0)}}{\sqrt{c_1(r_0)}}.
\end{equation}
By using Eqs.~(\ref{eq:cs})-(\ref{eq:ID}), $I_D$ in a strong deflection limit $r_0 \rightarrow r_\mathrm{m}$ or $b \rightarrow b_\mathrm{m}$ is obtained as
\begin{equation}
I_D =-\bar{a} \log \left( \frac{b}{b_\mathrm{m}}-1 \right) +\bar{a} \log 4,
\end{equation}
where $\bar{a}$ is given by
\begin{equation}
\bar{a}\equiv \frac{\sqrt{2}M}{\sqrt{4M^2-3Q^2}}.
\end{equation}

\begin{figure}[htbp]
\begin{center}
\includegraphics[width=85mm]{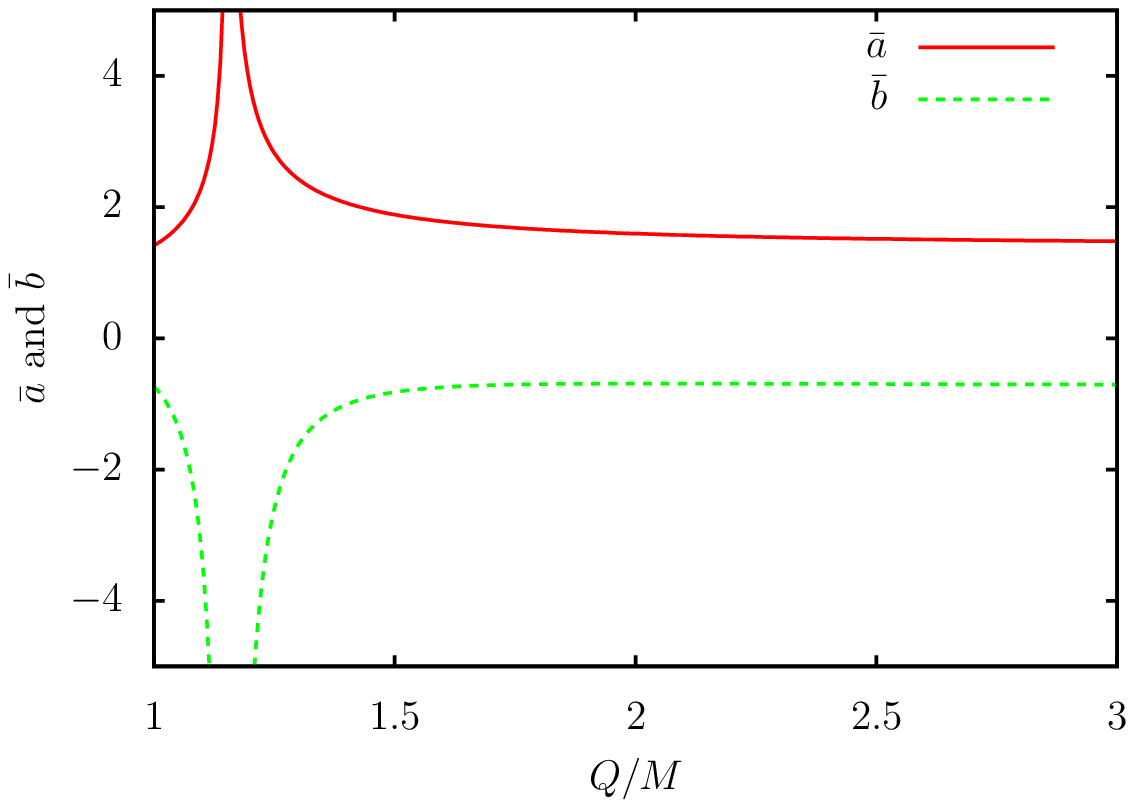}
\end{center}
\caption{ 
Solid (red) and dashed (green) curves denote the parameters $\bar{a}$ and $\bar{b}$ in the deflection angle, respectively.
}
\label{fig:ab}
\end{figure}

The regular part~$I_R$ is given by
\begin{equation}
I_R=\int^1_0 G(z,r_0)dz,
\end{equation}
where $G(z,r_0)$ is defined as
\begin{equation}
G(z,r_0)\equiv R(z,r_0) F(z,r_0) -R(0,r_\mathrm{m}) F_D(z,r_0).
\end{equation}
We expand $G(z,r_0)$ in the power of $r_0-r_\mathrm{m}$ as 
\begin{equation}
I_R(r_0)=\sum^\infty_{j=0} \frac{1}{j!} (r_0-r_\mathrm{m})^j \int^1_0 \left. \frac{\partial^j G}{\partial r_0^j} \right|_{r_0=r_\mathrm{m}} dz
\end{equation}
and we consider the first term, in which we are interested,
\begin{equation}
I_R=\int^1_0 G(z,r_\mathrm{m}) dz.
\end{equation}
We can obtain $\bar{b}$ as 
\begin{equation}
\bar{b} =\bar{a} \log 4 +I_R -\pi. 
\end{equation}
The parameters $\bar{a}$ and $\bar{b}$ are shown in Fig.~\ref{fig:ab}.
In the extreme charged Reissner-Nordstr\"{o}m  black hole limit $Q \rightarrow M$,
the parameters are obtained analytically as $\bar{a}=\sqrt{2}\sim 1.41$ and $\bar{b}=\sqrt{2}\log\left[ 32(3-2\sqrt{2}) \right] -\pi \sim -0.733$ 
as obtained in Refs.~\cite{Tsukamoto:2016jzh,Tsukamoto:2016oca}.

\subsubsection{For $Q/M>2/\sqrt{3}$}
In the case of $Q/M>2/\sqrt{3}$, there is a photon sphere with $b=b_\mathrm{m}=r_+^2/(r_+-M)$ on the throat at $r=r_\mathrm{m}=r_+$.
Since the factor $R(0,r_\mathrm{m})$ diverges at $z=0$, we rewrite $I$ as
\begin{equation}
I(r_0)=\int^{1}_0 S(z,r_0) H(z,r_0) dz,
\end{equation}
where $S(z,r_0)$ and $H(z,r_0)$ are defined by
\begin{equation}
S(z,r_0)\equiv 2\sqrt{M} r_0 \left[ r_0+M(z-1) \right]
\end{equation}
and 
\begin{equation}
H(z,r_0)\equiv \frac{1}{\sqrt{h(z,r_0)}},
\end{equation}
respectively, where $h(z,r_0)$ is given by
\begin{eqnarray}
h(z,r_0)&\equiv& \left\{(M-r_0)^2-(z-1)^2 \left[M (z-1)+r_0\right]^2\right\} \nonumber\\
&&\times \left\{M \left[Q^2 (z-1)^2+r_0^2\right]+2 Q^2 r_0 (z-1)\right\}. \nonumber\\
\end{eqnarray} 
Here, we have used $b(r_0)=r_0^2/(r_0-M)$.
We expand $h(z,r_0)$ around $z=0$ as
\begin{equation}
h(z,r_0)= \bar{c}_1(r_0)z + \bar{c}_2(r_0)z^2+ O(z^3),
\end{equation}
where $\bar{c}_1(r_0)$ and $\bar{c}_2(r_0)$ are given by 
\begin{eqnarray}
\bar{c}_1(r_0)\equiv 2 \left(2 M^2-3 M r_0+r_0^2\right) \left[ M \left(Q^2+r_0^2\right)-2 Q^2 r_0 \right] \nonumber\\
\end{eqnarray}
and
\begin{eqnarray}
\bar{c}_2(r_0)&\equiv& -14 M^3 Q^2+6 r_0^3 \left(M^2+Q^2\right) -M r_0^4 \nonumber\\
&&-M r_0^2 \left(6 M^2+29 Q^2\right)+38 M^2 Q^2 r_0, 
\end{eqnarray}
respectively.
Since we get $\bar{c}_{1\mathrm{m}}\equiv \bar{c}_1(r_\mathrm{m})=0$ 
and 
\begin{eqnarray}
&&\bar{c}_{2\mathrm{m}}\equiv \bar{c}_2(r_\mathrm{m}) =\frac{4 Q^2 \left(Q^2-M^2\right)}{M^3} \nonumber\\
&&\times \left[2 M^4-7 M^2 Q^2+(4 Q^2-5 M^2)Q \sqrt{Q^2-M^2} +4 Q^4\right], \nonumber\\
\end{eqnarray}
$H(z,r_0)$ diverges in order of $z^{-1}$ at $z=0$.
We expand $\bar{c}_{1}$ and $b$ around $r_0=r_\mathrm{m}$ as
\begin{eqnarray}\label{eq:cs2}
\bar{c}_{1}(r_0)= \bar{c}_{1\mathrm{m}}^\prime (r_0-r_\mathrm{m})+O\left(( r_0-r_\mathrm{m} )^2\right), 
\end{eqnarray}
where $\bar{c}_{1\mathrm{m}}^\prime$ is given by
\begin{eqnarray}
\bar{c}_{1\mathrm{m}}^\prime
&=&\frac{  2(M^2- Q^2) \sqrt{Q^2-M^2}+ 3 M^2 Q-2 Q^3}{M^2}\nonumber\\
&&\times 4 Q (M^2-Q^2),
\end{eqnarray}
and 
\begin{equation}\label{eq:bs2}
b(r_0)= b_\mathrm{m} + b^{\prime}_{\mathrm{m}} (r_0-r_\mathrm{m}) +O\left(( r_0-r_\mathrm{m} )^2\right), 
\end{equation}
where $b^{\prime}_{\mathrm{m}}$ is given by
\begin{equation}\label{eq:db}
b^{\prime}_{\mathrm{m}}= \frac{Q^2-2Q\sqrt{Q^2-M^2}}{M^2-Q^2},
\end{equation}
respectively, to use them later.

We separate $I$ into a divergent part $I_d$ and a regular part $I_r$.
The divergent part is defined by 
\begin{equation}
I_d\equiv \int^{1}_0 S(0,r_\mathrm{m}) H_d(z,r_0) dz,
\end{equation}
where $S(0,r_\mathrm{m})$ is obtained as
\begin{eqnarray}\label{eq:S02}
S(0,r_\mathrm{m})
=\frac{2 Q\sqrt{Q^2-M^2}\left( \sqrt{Q^2-M^2}+Q \right)^2 }{M^{3/2}}
\end{eqnarray}
and $H_d(z,r_0)$ is defined by
\begin{equation}\label{eq:HD}
H_d(z,r_0)\equiv \frac{1}{\sqrt{\bar{c}_1(r_0) z+ \bar{c}_2(r_0) z^2}}.
\end{equation}
The divergent term $I_d$ can be integrated as~\cite{Bozza:2002zj,Tsukamoto:2016jzh}
\begin{equation}\label{eq:ID2}
I_d = \frac{2S(0,r_\mathrm{m})}{\sqrt{\bar{c}_2(r_0)}} \log \frac{\sqrt{\bar{c}_2(r_0)}+\sqrt{\bar{c}_1(r_0)+\bar{c}_2(r_0)}}{\sqrt{\bar{c}_1(r_0)}}.
\end{equation}
From Eqs.~(\ref{eq:cs2})-(\ref{eq:ID2}), $I_d$ in a strong deflection limit $r_0 \rightarrow r_\mathrm{m}$ or $b \rightarrow b_\mathrm{m}$ is given by 
\begin{equation}
I_d =-\bar{a} \log \left( \frac{b}{b_\mathrm{m}}-1 \right) +\bar{a} \log \frac{8\sqrt{Q^2-M^2}-4Q}{\sqrt{Q^2-M^2}},
\end{equation}
where $\bar{a}$ is obtained as
\begin{eqnarray}
\bar{a}=\frac{(Q + \sqrt{Q^2-M^2})^2}{\sqrt{2M^4-7M^2Q^2+4Q^4+(4Q^2-5M^2)Q\sqrt{Q^2-M^2} }}. \nonumber\\
\end{eqnarray}

We define the regular part~$I_r$ as
\begin{equation}
I_r\equiv \int^1_0 g(z,r_0)dz,
\end{equation}
where $g(z,r_0)$ is given by
\begin{equation}
g(z,r_0)\equiv S(z,r_0) H(z,r_0) -S(0,r_\mathrm{m}) H_d(z,r_0)
\end{equation}
and it can be expanded in the power of $r_0-r_\mathrm{m}$ as 
\begin{equation}
I_r(r_0)=\sum^\infty_{j=0} \frac{1}{j!} (r_0-r_\mathrm{m})^j \int^1_0 \left. \frac{\partial^j g}{\partial r_0^j} \right|_{r_0=r_\mathrm{m}} dz.
\end{equation}
The first term in which we are interested is given by
\begin{equation}
I_r=\int^1_0 g(z,r_\mathrm{m}) dz.
\end{equation}
Thus, we can express $\bar{b}$ as 
\begin{equation}
\bar{b} =\bar{a} \log \frac{8\sqrt{Q^2-M^2}-4Q}{\sqrt{Q^2-M^2}}+I_r -\pi.
\end{equation}
Notice that $\bar{a}$ and $\bar{b}$ are shown in Fig.~\ref{fig:ab}.
The ultrastatic limit $M \rightarrow 0$ under a fixed ADM mass $m$,
the deflection angle of the ray has
$\bar{a}=\sqrt{2}\sim 1.41$ and $\bar{b}=2\sqrt{2}\log\left[ 4(2-\sqrt{2}) \right] -\pi=\sqrt{2}\log\left[ 32(3-2\sqrt{2}) \right] -\pi \sim -0.733$ 
as obtained by Tsukamoto and Harada~\cite{Tsukamoto:2016zdu}. 
We notice that the parameters $\bar{a}$ and $\bar{b}$ are coincident with the ones in the case of $Q\rightarrow M$.

\subsection{Observables in the strong deflection limit}
We introduce an angle $\theta^0_N$ defined by
\begin{equation}\label{eq:theta0n}
\alpha(\theta^0_N) = 2\pi N
\end{equation}
and we expand the deflection angle $\alpha(\theta)$ around $\theta=\theta^0_N$ to obtain
\begin{equation}\label{eq:alphaexpand}
\alpha(\theta)=\alpha(\theta^0_N)+\left. \frac{d\alpha}{d\theta} \right|_{\theta=\theta^0_N} (\theta-\theta^0_N)+O\left( (\theta-\theta^0_N)^2 \right).
\end{equation}
In the strong deflection limit $b\rightarrow b_{\mathrm{m}}+0$, 
the deflection angle is expressed by
\begin{eqnarray}\label{eq:alphaSDL}
\alpha(\theta)
&=&-\bar{a} \log \left( \frac{\theta}{\theta_\infty}-1 \right)+\bar{b} \nonumber\\
&&+O\left( \left( \frac{\theta}{\theta_\infty}-1 \right) \log \left( \frac{\theta}{\theta_\infty}-1 \right) \right),
\end{eqnarray}
where $\theta_\infty \equiv b_{\mathrm{m}}/D_{\mathrm{ol}}$ is the image angle of the photon sphere, and 
$\theta^0_N$ is obtained as, from Eqs. (\ref{eq:theta0n}) and (\ref{eq:alphaSDL}), 
\begin{equation}\label{eq:theta0ninf}
\theta^0_N=\left( 1+e^{\frac{\bar{b}-2\pi N}{\bar{a}}} \right) \theta_\infty
\end{equation}
and we obtain, from Eq. (\ref{eq:alphaSDL}), 
\begin{equation}\label{eq:dalpdthe}
\left. \frac{d\alpha}{d\theta} \right|_{\theta=\theta^0_N}=\frac{\bar{a}}{\theta_\infty-\theta^0_N}.
\end{equation}
The effective deflection angle for the positive solution $\theta=\theta_N$ of the lens equation with a positive winding number $N$ is given by, 
from Eqs. (\ref{eq:defn}), (\ref{eq:theta0n}), (\ref{eq:alphaexpand}), (\ref{eq:theta0ninf}), and (\ref{eq:dalpdthe}),
\begin{equation}\label{eq:althn}
\bar{\alpha}(\theta_N)=\frac{\bar{a}}{\theta_\infty e^{\frac{\bar{b}-2\pi N}{\bar{a}}}} ( \theta^0_N-\theta_N ).
\end{equation}
By substituting the effective deflection angle (\ref{eq:althn}) into the lens equation (\ref{eq:lens}), we obtain the positive solution of the lens equation 
or the image angle for the positive winding number $N$ as
\begin{equation}
\theta_N(\phi) \sim \theta^0_N - \frac{\theta_\infty e^{\frac{\bar{b}-2\pi N}{\bar{a}}} D_\mathrm{os} (\theta^0_N-\phi)}{\bar{a}D_\mathrm{ls}},
\end{equation}
the radius of an Einstein ring with the positive winding number $N$ as 
\begin{equation}
\theta_{\mathrm{E}N}\equiv \theta_N(0) \sim \left( 1- \frac{\theta_\infty e^{\frac{\bar{b}-2\pi N}{\bar{a}}}D_\mathrm{os}}{\bar{a}D_\mathrm{ls}} \right) \theta^0_N,
\end{equation}
and the difference between the image angle with $N=1$ and the innermost image angle $\bar{\mathrm{s}}$ as
\begin{equation}
\bar{\mathrm{s}}\equiv \theta_1-\theta_\infty \sim \theta^0_1-\theta^0_\infty = \theta_\infty e^{\frac{\bar{b}-2\pi}{\bar{a}}}.
\end{equation}
The magnification of the image with $\theta_N(\phi)$ for each positive winding number $N$ is given by
\begin{equation}
\mu_N \equiv \frac{\theta_N}{\phi}\frac{d\theta_N}{d\phi} \sim \frac{\theta^2_\infty D_{\mathrm{os}} \left(1+ e^{\frac{\bar{b}-2\pi N}{\bar{a}}}\right) e^{\frac{\bar{b}-2\pi N}{\bar{a}}}}{\phi \bar{a} D_{\mathrm{ls}}},
\end{equation}
the sum of the magnifications from $N=1$ to $\infty$ is given by
\begin{eqnarray}
\sum^\infty_{N=1} \mu_N \sim \frac{\theta^2_\infty D_{\mathrm{os}} \left(1+e^{\frac{2\pi}{\bar{a}}} +e^{\frac{\bar{b}}{\bar{a}}}\right) e^{\frac{\bar{b}}{\bar{a}}}}{\phi \bar{a} D_{\mathrm{ls}}\left( e^{\frac{4\pi}{\bar{a}}} -1 \right)},
\end{eqnarray}
and the ratio of the magnification of the image with $N=1$ to the others $\bar{\mathrm{r}}$ is given by
\begin{equation}
\bar{\mathrm{r}} 
\equiv \frac{\mu_1}{\sum^\infty_{N=2} \mu_N}
\sim \frac{\left( e^{\frac{4\pi}{\bar{a}}}-1 \right) \left( e^{\frac{2\pi}{\bar{a}}}+e^{\frac{\bar{b}}{\bar{a}}} \right)}{ e^{\frac{2\pi}{\bar{a}}}+e^{\frac{4\pi}{\bar{a}}} +e^{\frac{\bar{b}}{\bar{a}}} },
\end{equation}
where $\sum^\infty_{N=2} \mu_N$ is  
\begin{eqnarray}
\sum^\infty_{N=2} \mu_N \sim \frac{\theta^2_\infty D_{\mathrm{os}} \left(e^{\frac{2\pi}{\bar{a}}}+e^{\frac{4\pi}{\bar{a}}} +e^{\frac{\bar{b}}{\bar{a}}}\right) e^{\frac{\bar{b}-4\pi}{\bar{a}}}}{\phi \bar{a} D_{\mathrm{ls}}\left( e^{\frac{4\pi}{\bar{a}}} -1 \right)}.
\end{eqnarray}

\section{Schwarzschild spacetime}
We consider a Schwarzschild spacetime to compare observables with ones in the Bronnikov-Kim spacetime.
The line element in the Schwarzschild spacetime is given by 
\begin{equation}
ds^{2}
=-\left(1-\frac{2m_\mathrm{S}}{r}\right)dt^{2} +\frac{dr^{2}}{1-\frac{2m_\mathrm{S}}{r}}+r^{2}(d\vartheta^{2}+\sin^{2}\vartheta d\varphi^{2}),
\end{equation}
where $m_\mathrm{S}$ is its ADM mass.

Under the weak-field approximation, the line element can be expressed by
\begin{eqnarray}
ds^{2}
&=&-\left(1-\frac{2m_\mathrm{S}}{r}\right)dt^{2} +\left(1+\frac{2m_\mathrm{S}}{r}\right)dr^{2} \nonumber\\
&&+r^{2}(d\vartheta^{2}+\sin^{2}\vartheta d\varphi^{2}).
\end{eqnarray}
If we read the parameters $Q$, $M$, and $m$ in the Bronnikov-Kim spacetime as $m_\mathrm{S}$, 
we obtain the deflection angle and the observables under the weak-field approximation in the Schwarzschild spacetime. 
The deflection angle of a ray under the weak-field approximation is given by
\begin{eqnarray}\label{eq:weakdef}
\alpha \sim \frac{4m_\mathrm{S}}{b}
\end{eqnarray}
and the radius of the Einstein ring is given by
\begin{equation}
\theta_{\mathrm{E}0} = \sqrt{\frac{4m_\mathrm{S}D_{\mathrm{ls}}}{D_{\mathrm{os}}D_{\mathrm{ol}}}}.
\end{equation}

In the strong deflection limit $b\rightarrow b_\mathrm{m}=3\sqrt{3}m_\mathrm{S}$, the parameters $\bar{a}$ and $\bar{b}$ in the deflection angle are obtained as
$\bar{a}=1$ and $\bar{b}=\log[ 216(7-4\sqrt{3}) ]-\pi\sim -0.40$, respectively~\cite{Bozza_Capozziello_Iovane_Scarpetta_2001,Bozza:2002zj}.

\section{Discussion and Conclusion}
We have investigated the gravitational lensing by the Bronnikov-Kim wormhole under the weak-field approximation 
and in the strong deflection limit. 
The Bronnikov-Kim wormhole metric is the same as 
the one of a wormhole suggested in Einstein-Dirac-Maxwell theory~\footnote{A thin shell at the throat is discussed in Refs.~\cite{Bolokhov:2021fil,Konoplya:2021hsm,Blazquez-Salcedo:2021udn,Danielson:2021aor}.}
in a simple case~\cite{Blazquez-Salcedo:2020czn,Bolokhov:2021fil}.
The metric becomes the one of an extreme charged Reissner-Nordstr\"{o}m black hole in a limit $Q\rightarrow M$
and the one of a spatial Schwarzschild wormhole~\cite{Dadhich:2001fu} 
in an ultrastatic limit $M\rightarrow0$ under a fixed ADM mass $m$.
The parameter $\bar{b}$ of the Bronnikov-Kim wormhole has been calculated numerically partly while
$\bar{b}$ of the extreme charged Reissner-Nordstr\"{o}m black hole~\cite{Tsukamoto:2016jzh,Tsukamoto:2016oca} 
and the spatial Schwarzschild wormhole~\cite{Tsukamoto:2016zdu}
are obtained analytically.
Interestingly, in both cases of the extreme charged Reissner-Nordstr\"{o}m black hole and the spatial Schwarzschild wormhole, 
we obtain exactly the same parameters $\bar{a}=\sqrt{2}$ 
and $\bar{b}=\sqrt{2}\log\left[ 32(3-2\sqrt{2}) \right] -\pi$. 

Recently, in Ref.~\cite{Jusufi:2021lei}, Jusufi~\textit{et al.} have discussed shadow images under an assumption 
that a supermassive compact object at the center of our Galaxy is the Bronnikov-Kim wormhole. 
From observational data on the orbit of S2 star~\cite{Do:2019txf}, 
they have concluded that the parameters of the metric are $Q\sim 8\times 10^6M_\odot$ and $M\sim 4\times 10^6M_\odot$.
Note that a Schwarzschild black hole with an ADM mass $m_\mathrm{S}=4\times 10^6 M_{\odot}$, which matches the observation of orbit of S2 star,  
can form an Einstein ring with a diameter $2\theta_{\mathrm{E}0}\sim 2.86$ arcsecond and a photon sphere with a diameter $2\theta_{\infty}\sim 51.58$ $\mu$as
if we set distances $D_{\mathrm{os}}=16$ and $D_{\mathrm{ol}}=D_{\mathrm{ls}}=8$~kpc.
On the other hand, the Bronnikov-Kim wormhole with the parameters $Q= 8\times 10^6M_\odot$ and $M= 4\times 10^6M_\odot$
has $2\theta_{\mathrm{E}0}\sim 4.52$ arcsecond and $2\theta_{\infty}\sim 85.56$ $\mu$as as shown in Table~I.
Therefore, we would distinguish the Schwarzschild black hole with ADM mass $m_\mathrm{S}=4\times 10^6 M_{\odot}$
from the Bronnikov-Kim wormhole with the parameters $Q\sim 8\times 10^6M_\odot$ and $M\sim 4\times 10^6M_\odot$ 
at the center of our Galaxy 
if lensed images under the weak-field approximation and $D_{\mathrm{os}}$ are observed or if lensed images in the strong deflection limit are observed.
\begin{table*}[htbp]
 \caption{Parameters~$\bar{a}$ and $\bar{b}$,  
 the diameters of the photon sphere~$2\theta_{\infty}$ and the outermost image~$2\theta_{\mathrm{E}1}$, 
 the difference of their radii $\bar{\mathrm{s}}=\theta_{\mathrm{E}1}-\theta_\infty$, 
 the magnification of the pair of the outermost images $\mu_{1\mathrm{tot}}(\phi) \sim 2 \left| \mu_{1} \right|$, 
 and the ratio of the magnification of the outermost image to the others $\bar{\mathrm{r}}= \mu_1/\sum^\infty_{N=2} \mu_N$ 
 in the strong deflection limit for given $Q/M$ are shown.
 The diameter of an Einstein ring $2\theta_{\mathrm{E}0}$ under the weak-field approximation is also shown.
  As a reference, the ones of the Schwarzschild spacetime also are considered.
 We set the parameters $M=m_\mathrm{S}=4\times 10^6 M_{\odot}$, distances $D_{\mathrm{os}}=16$ and $D_{\mathrm{ol}}=D_{\mathrm{ls}}=8$~kpc, 
 and the source angle $\phi=1$ arcsecond. 
 }
\begin{center}
\begin{tabular}{ c c c c c c c c c c c |c} \hline\hline
$Q/M$          	                   	 &&$1$        &$1.1$    &$1.2$     &$1.5$     &$1.8$    &$2$      &$2.2$    &$2.5$    &$3$      &Schwarzschild \\ \hline
$\bar{a}$        	            	 &&$1.41$     &$2.33$   &$3.84$    &$1.89$    &$1.66$   &$1.60$   &$1.56$   &$1.52$   &$1.48$   &$1.00$ \\ 
$\bar{b}$        	            	 &&$-0.733$   &$-3.312$ &$-6.150$  &$-0.820$  &$-0.698$ &$-0.689$ &$-0.690$ &$-0.696$ &$-0.705$ &$-0.400$ \\ 
$2\theta_{\infty}$~($\mu$as)	         &&$39.71$    &$39.71$  &$40.15$   &$52.30$   &$70.84$  &$85.56$  &$102.0$  &$129.7$  &$184.1$  &$51.58$ \\ 
$2\theta_{\mathrm{E}1}$~($\mu$as)        &&$40.00$    &$40.35$  &$41.72$   &$53.51$   &$71.91$  &$86.64$  &$103.1$  &$131.0$  &$185.7$  &$51.65$ \\ 
$\bar{\mathrm{s}}$~($\mu$as)             &&$0.14$     &$0.32$   &$0.76$    &$0.60$    &$0.53$   &$0.54$   &$0.57$   &$0.61$   &$0.82$   &$0.03$ \\ 
$\mu_{1\mathrm{tot}}(\phi)\times10^{17}$ &&$3.8$      &$5.4$    &$8.3$     &$17$      &$22$     &$29$     &$37$     &$55$     &$100$    &$1.6$ \\ 
$\bar{\mathrm{r}}$                	 &&$85$       &$14$     &$4.3$     &$28$      &$43$     &$51$     &$56$     &$63$     &$69$     &$535$ \\ 
$2\theta_{\mathrm{E}0}$~(arcsecond)      &&$2.86$     &$3.01$   &$3.16$    &$3.65$    &$4.17$   &$4.52$   &$4.89$   &$5.45$   &$6.40$   &$2.86$ \\ 
\hline\hline
\end{tabular}
\end{center}
\end{table*}

The qualitative features of the gravitational lensing by the Bronnikov-Kim wormhole in the strong deflection limit are the same as the ones 
in the Schwarzschild spacetime because both cases follow general features of lensing by a photon sphere.
Distinguishing the Bronnikov-Kim wormhole from the Schwarzschild black hole with only the images with winding numbers $N=1$ and $N\geq 2$ 
is a challenging future work
since the difference of their observables in the strong deflection limit is small.
The ratio of the magnifications of the image with $N=1$ to the others $\bar{r}$ and
$\bar{\mathrm{s}}/\theta_{\infty} \sim  e^{(\bar{b}-2\pi)/\bar{a}}$
can be used for careful verifications of the spacetimes in the future since 
they are not affected by the details of the lens configuration.
Table~II shows the observables of a supermassive object at the center of galaxy M87 with the parameters $m_\mathrm{S}=6.5\times 10^9 M_{\odot}$ 
and $D_{\mathrm{ol}}=16.8$~Mpc estimated in Ref.~\cite{Akiyama:2019cqa}. 
We set the parameters $M$ and $Q$ of the Bronnikov-Kim wormhole so that $\theta_{\infty}$ of the wormhole is the same value 
as $\theta_{\infty}$ of the Schwarzschild black hole by using
$4M=3\sqrt{3}m_\mathrm{S}$
for $Q/M<2/\sqrt{3}\sim 1.15$ and 
$r_+^2/(r_+ -M)=3\sqrt{3}m_\mathrm{S}$
for 
$Q/M>2/\sqrt{3}$.
\begin{table*}[htbp]
 \caption{In the case of the parameters $m_\mathrm{S}=6.5\times 10^9 M_{\odot}$, distances $D_{\mathrm{os}}=33.6$ 
 and $D_{\mathrm{ol}}=D_{\mathrm{ls}}=16.8$~Mpc, 
 and the source angle $\phi=1$ arcsecond is shown. 
 The parameters $M$ and $Q$ are set so that $\theta_{\infty}$ of the Bronnikov-Kim wormhole is the same value 
 as $\theta_{\infty}$ of the Schwarzschild black hole. 
 }
\begin{center}
\begin{tabular}{ c c c c c c c c c c c |c} \hline\hline
$Q/M$          	                   	 &&$1$        &$1.1$    &$1.2$     &$1.5$     &$1.8$    &$2$      &$2.2$    &$2.5$    &$3$      &Schwarzschild \\ \hline
$\bar{a}$        	            	 &&$1.41$     &$2.33$   &$3.84$    &$1.89$    &$1.66$   &$1.60$   &$1.56$   &$1.52$   &$1.48$   &$1.00$ \\ 
$\bar{b}$        	            	 &&$-0.733$   &$-3.312$ &$-6.150$  &$-0.820$  &$-0.698$ &$-0.689$ &$-0.690$ &$-0.696$ &$-0.705$ &$-0.400$ \\ 
$2\theta_{\infty}$~($\mu$as)	         &&$39.91$    &$39.91$  &$39.91$   &$39.91$   &$39.91$  &$39.91$  &$39.91$  &$39.91$  &$39.91$  &$39.91$ \\ 
$2\theta_{\mathrm{E}1}$~($\mu$as)        &&$40.19$    &$40.56$  &$41.47$   &$40.84$   &$40.51$  &$40.42$  &$40.36$  &$40.31$  &$40.27$  &$39.96$ \\ 
$\bar{\mathrm{s}}$~($\mu$as)             &&$0.14$     &$0.32$   &$0.78$    &$0.46$    &$0.30$   &$0.25$   &$0.23$   &$0.20$   &$0.18$   &$0.025$ \\ 
$\mu_{1\mathrm{tot}}(\phi)\times10^{17}$ &&$3.9$      &$5.4$    &$8.2$     &$9.7$     &$7.1$    &$6.2$    &$5.7$    &$5.2$    &$4.7$    &$0.97$ \\ 
$\bar{\mathrm{r}}$                	 &&$85$       &$14$     &$4.3$     &$28$      &$43$     &$51$     &$56$     &$63$     &$69$     &$535$ \\ 
$2\theta_{\mathrm{E}0}$~(arcsecond)      &&$2.87$     &$3.02$   &$3.15$    &$3.19$    &$3.13$   &$3.09$   &$3.06$   &$3.02$   &$2.98$   &$2.52$ \\ 
\hline\hline
\end{tabular}
\end{center}
\end{table*}

On this paper, we do not treat the gravitational lensing by the marginally unstable photon sphere on the throat with $Q/M=2/\sqrt{3}$ 
since the deflection angle would diverge nonlogarithmically as discussed in Ref.~\cite{Tsukamoto:2020uay}.

%
\appendix
\section{Arnowitt-Deser-Misner (ADM) mass}
We calculate an ADM mass~\cite{Poisson} of the Bronnikov-Kim wormhole.
Under the weak-field approximation, the line element can be expressed as
\begin{eqnarray}
ds^{2}
&\sim&-\left(1-\frac{2M}{r}\right) dt^{2} \nonumber\\
&&+\left(1+\frac{2Q^2}{Mr}\right)dr^{2}+r^{2}(d\vartheta^{2}+\sin^{2}\vartheta d\varphi^{2}) \nonumber\\
&=&-\left(1-\frac{2M}{r(r_*)}\right) dt^{2} \nonumber\\
&&+\left(1+\frac{2Q^2}{Mr_*}\right) \left[ dr_*^{2}+r_*^2(d\vartheta^{2}+\sin^{2}\vartheta d\varphi^{2}) \right].
\end{eqnarray}
We consider a hypersurface $\Sigma_t$ which is the surface of constant $t$.
A unit normal to the hypersurface $\Sigma_t$ is given by $n_\mu=-(1-M/r)\partial_t$
and an induced metric on the hypersurface is given by
\begin{equation}
h_{ab} dy^a dy^b = \left(1+\frac{2Q^2}{Mr_*}\right) \left[ dr_*^{2}+r_*^2(d\vartheta^{2}+\sin^{2}\vartheta d\varphi^{2}) \right].
\end{equation}
Its boundary $S_t$ is a two-sphere $r_*=R_*$ and its unit normal is given by 
\begin{equation}
r_{*a}=\left(1+\frac{Q^2}{Mr_*}\right) \partial_{r_{*}}
\end{equation}
and an induced metric on the boundary $S_t$ is given by
\begin{equation}
\sigma_{AB} d\theta^A d\theta^B = \left(1+\frac{2Q^2}{MR_*}\right) R_*^2(d\vartheta^{2}+\sin^{2}\vartheta d\varphi^{2}).
\end{equation}
The ADM mass $m$ is obtained as
\begin{eqnarray}
m\equiv-\frac{1}{8\pi} \lim_{S_{\mathrm{t}}\rightarrow \infty} \oint_{S_{\mathrm{t}}}(k-k_{0})\sqrt{\sigma}d^{2}\theta 
=\frac{Q^2}{M},
\end{eqnarray}
where the extrinsic curvature of $S_t$ embedded in $\Sigma_t$ is given by
\begin{eqnarray}
k=r^a_{*|a}=\frac{2}{R_*} \left( 1- \frac{2Q^2}{MR_*} \right),
\end{eqnarray}
the extrinsic curvature of $S_t$ embedded in a flat space is given by
\begin{eqnarray}
k_0=\frac{2}{\left(1+\frac{2Q^2}{MR_*}\right)^{1/2} R_*}=\frac{2}{R_*} \left( 1- \frac{Q^2}{MR_*} \right),
\end{eqnarray}
and the leading term of $\sigma$ is given by $\sigma=R_*^4 \sin^{2}\vartheta$.

\end{document}